\documentclass[showpacs]{revtex4}
\usepackage{graphicx}

\begin{document}

\title{Weak-coupling functional renormalization-group analysis of the 
Hubbard model on the anisotropic triangular lattice} 

\author{Shan-Wen Tsai\footnote{new address: Institute for Fundamental 
Theory and Department of Physics, University of 
Florida, Gainesville, FL 32611-8440 (tsai@phys.ufl.edu)} and J. B. Marston}

\affiliation{Department of Physics, Brown University, Providence, RI 02912-1843}

\date{\today}  

\begin{abstract}
Motivated by experiments on the layered compounds $\kappa$-(BEDT-TTF)$_2$X, 
Cs$_2$CuCl$_4$, and very recently Na$_x$CoO$_2 \cdot y$H$_2$O, 
we present a weak-coupling functional renormalization-group analysis of the Hubbard model on the 
anisotropic triangular lattice.  As the model interpolates between the
nearest-neighbor square lattice and decoupled chains via the isotropic
triangular lattice, it permits the study of competition between antiferromagnetic and BCS Cooper
instabilities.  We begin by reproducing known results for decoupled chains, 
and for the square lattice with only nearest-neighbor hopping amplitude $t_1$. 
We examine both repulsive and attractive Hubbard interactions.  
The role of formally irrelevant contributions to the one-loop
renormalization-group flows is also studied, and these subleading contributions
are shown to be important in some instances.  We then 
observe that crossover to a BCS-dominated regime can 
occur even at half-filling when antiferromagnetism is frustrated through the introduction of 
a next-nearest-neighbor hopping amplitude $t_2$ along one of the two diagonal directions.  
Stripes are not expected to occur and time-reversal breaking 
$d_{x^2 - y^2} \pm i d_{xy}$ superconducting order does not arise spontaneously; 
instead pure $d_{x^2 - y^2}$ order is favored.  At the isotropic triangular point 
($t_1 = t_2$) we find the possibility of re-entrant antiferromagnetic long-range order.
\end{abstract}


\maketitle

\section{Introduction}
\label{sec:Intro}

The behavior of strongly correlated electrons moving in reduced spatial
dimensions continues to yield surprising new physics. For example,   
intriguing experiments\cite{Williams,Ishiguro,McKenzie1} on the 
$\kappa$-(BEDT-TTF)$_2$X family of layered organic molecular crystals 
evoke similar findings in the field of high-temperature cuprate 
superconductivity.  Like the high-T$_c$ cuprates, the layered organic materials 
exhibit a wide variety of electronic properties.  In particular,
the phase diagram is rather similar to that of the 
cuprates\cite{McKenzie2} and there is some evidence for unconventional pairing 
with nodes in the gap from NMR relaxation rate\cite{nmr1,nmr2,nmr3}, 
specific heat\cite{specificheat}, penetration 
depth\cite{pdepth1,pdepth2,pdepth3,pdepth4,pdepth5},
STM spectroscopy\cite{stm}, mm-wave transmission\cite{mm1} 
(see however Refs. \onlinecite{mm2,mm3}) and thermal conductivity\cite{thermal1,thermal2} 
measurements.  Other experiments, however, suggest 
$s$-wave pairing\cite{swave1,swave2,swave3,swave4,swave5,swave6}.  
Competition between 
antiferromagnetic and superconducting instabilities, seen in the cuprates, 
also seems to occur in the $\kappa$-(BEDT-TTF)$_2$X compounds\cite{Lefebvre}.  
In contrast to the square CuO$_2$ lattice of the high-temperature
superconductors, the organic molecules pair up into dimers, and these dimers 
form a triangular lattice.  

Two other quasi two-dimensional materials with triangular lattices have been
the subject of recent attention: The antiferromagnetic insulator 
Cs$_2$CuCl$_4$ compound\cite{Radu} and the cobalt-based superconductor
Na$_x$CoO$_2 \cdot y$H$_2$O that may be an analog of the cuprate
high-temperature superconductors\cite{Takada}.  Neutron scattering experiments
suggest the existence of deconfined spinon (spin-$1/2$) excitations in the
Cs$_2$CuCl$_4$ once antiferromagnetic order has been eliminated by heating 
the sample to the relatively low temperature of approximately 0.6K, or upon 
application of a field parallel to the plane\cite{Radu}. 
Geometric frustration of the spin-spin interactions is likely responsible
for the observed spin-liquid behavior.  As the cobalt atoms in the
Na$_x$CoO$_2 \cdot y$H$_2$O material also form a triangular 
lattice\cite{Balsys}, and as they have further been argued to carry spin-$1/2$ 
moments\cite{Takada}, we tentatively group this system into the same 
category as $\kappa$-(BEDT-TTF)$_2$X and Cs$_2$CuCl$_4$. 

\begin{figure}
\includegraphics[width=1.5in]{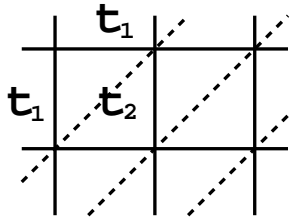}
\caption{Anisotropic triangular lattice with two hopping amplitudes $t_1$ 
and $t_2$.  The limit $t_2 = 0.0$ corresponds to the usual nearest-neighbor 
square lattice and $t_1 = 0.0$ corresponds to decoupled chains.} 
\label{fig:tri}
\end{figure}

Clearly theoretical investigations of strongly correlated electrons 
on triangular lattices are of great interest.  Initial studies
of strongly correlated systems often start with a minimal Hubbard model,
leaving extensions such as the inclusion of long-range Coulomb interactions 
for later more detailed work.  In fact
McKenzie has proposed\cite{McKenzie2} that a Hubbard model on the anisotropic 
triangular lattice serves as a minimal model of the conducting layers of 
$\kappa$-(BEDT-TTF)$_2$X.  It represents a simplification of a model 
introduced earlier by Kino and Fukuyama\cite{Kino1}.  Two distinct hopping 
matrix elements are introduced and the Hamiltonian is defined by:
\begin{eqnarray}
H = -t_1 \sum_{<{\bf ij}>} (c_{\bf i}^{\dagger \sigma} c_{{\bf j} \sigma} + 
H.c.) - t_2 \sum_{<<{\bf ij} >>} (c_{\bf i}^{\dagger \sigma} c_{{\bf j} 
\sigma} + H.c.) + ~U \sum_{\bf i} n_{{\bf i} \uparrow} n_{{\bf 
i} \downarrow} - \mu \sum_{\bf i} n_{\bf i},
\label{eq:Hamiltonian}
\end{eqnarray}
where $<{\bf ij}>$ denotes nearest-neighbor pairs of sites on the square 
lattice and $<<{\bf ij}>>$ denotes next-nearest-neighbor pairs along one of 
the two diagonal directions of the square lattice as shown in Fig. \ref{fig:tri}.
Quantum chemistry calculations suggest that, unlike the cuprate materials, 
in the case of the organic 
$\kappa$-(BEDT-TTF)$_2$X compounds the Hubbard interaction $U \approx t$. 
Thus a weak-coupling renormalization-group (RG) approach such as we adopt here 
may be expected to be reasonably accurate for the organic materials. 

\begin{figure}[!ht]
\includegraphics[width=5in]{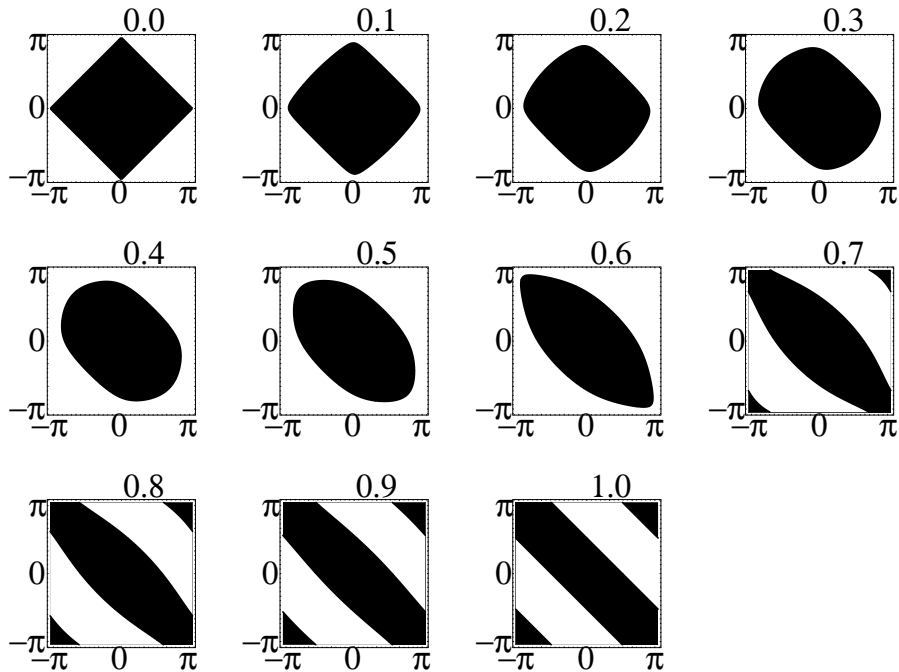}
\caption{Fermi surface of non-interacting electrons for different ratios of 
the two hopping amplitudes.  The number at the top of each graph is the 
anisotropy ratio $t_2/(t_1 + t_2)$ which ranges from $0$ (square lattice) 
to $1$ (decoupled chains).  The chemical potential $\mu$ is varied to 
ensure that the system remains half-filled.}
\label{fig:fs}
\end{figure}

The model is also interesting in its own right as it interpolates between 
the square lattice and decoupled chains. At half-filling, the non-interacting Fermi surface 
is perfectly nested in these two extreme limits. 
As nesting is imperfect in between the limiting cases, several 
phase transitions can be expected.  The square lattice, which has been the subject 
of many studies, corresponds to the special case of zero next-nearest-neighbor 
hopping, $t_2 = 0$.   When the repulsive interaction is turned on, 
nesting induces a spin density wave instability.  In the opposite limit, $t_1 = 0$,
the chains are completely decoupled.  These isolated chains of course have no spin order and
are described by the exact Bethe ansatz solution of Lieb and Wu\cite{Lieb}.
We pay particular attention to the intermediate region of
$t_1 \neq 0$ and $t_2 \neq 0$ and study it via a weak-coupling 
renormalization-group analysis.  The special isotropic triangular lattice point 
corresponds to $t_1 = t_2$.  Values for the hopping matrix elements obtained from
experiments and from quantum chemistry calculations for  
the conducting layer of $\kappa$-(BEDT-TTF)$_2$X suggest $t_1 > t_2$, that is,
somewhere intermediate between the square and the isotropic triangular limits.
The lattice anisotropy can be altered by uniaxial stress applied along the 
principal axes of the quasi-two-dimensional organic compound\cite{Choi}.  Fermi 
surfaces of non-interacting electrons for different ratios of the 
hopping matrix elements are shown in Fig. \ref{fig:fs}.  

In the next section we briefly introduce the RG
method we employ, a method first
implemented by Zanchi and Schulz\cite{Zanchi} for the case of the square lattice.  
We then present results of our calculations at different values of the anisotropy: 
decoupled chains (studied as a test case to check the reliability of the 
calculation), square lattice, and finally the anisotropic region intermediate
between the square lattice and the isotropic triangular lattice.  
We discuss the ordering tendencies and work out the implied phase 
diagram as a function of anisotropy parameter $t_2/(t_1 + t_2)$ which ranges from 
$0$ (square lattice) to $1$ (decoupled chains).  We also make comparison to
results obtained via other methods in the strong-coupling limit of 
large on-site repulsion.
 
\section{Renormalization-group calculation}
\label{sec:rg}

We follow the weak-coupling renormalization-group analysis
implemented by Zanchi and Schulz\cite{Zanchi} for interacting 
fermions on a two-dimensional lattice.  Like some previous work\cite{HKM}, 
the approach generalizes Shankar's renormalization group theory\cite{Shankar} 
to Fermi surfaces of arbitrary shape. 
More significantly, in principle the {\it only} approximation that is made in the 
approach of Zanchi and Schulz is an expansion in powers of 
the interaction strength, the on-site Coulomb interaction $U$.  Subleading terms 
generated during the RG transformations, which are dropped as irrelevant in the simplest
versions of the RG, are instead kept in this formulation.  Specifically, the
formally irrelevant, non-logarithmic, terms that appear in the six-point function 
during the process of mode elimination do in fact contribute to the RG flows.  
Thus while the simplest weak-coupling RG analyses makes a double expansion in both
the interaction strength and in the relevance of the terms retained in the renormalization
flows, in the approach of Zanchi and Schulz there is 
only a single expansion in the interaction strength.  (In practice some additional
approximations are made for computational convenience, as detailed below.  
These simplifications are not expected to alter the results significantly.) 
As we show below, in some cases this more accurate treatment leads to 
substantial differences in the RG flows. 

\begin{figure}[!ht]
\includegraphics[width=3in]{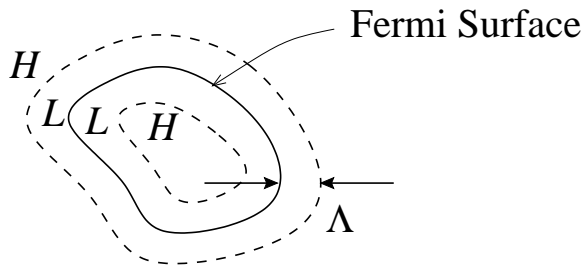}
\caption{Mode elimination is carried out in infinitesimal steps.  The figure 
shows the low energy modes (denoted $L$) which are inside a shell of thickness 
$\Lambda$ around the Fermi surface, and the high energy modes ($H$), 
which have already been integrated out and which are now outside the 
shell.  At each step the 
on-shell modes precisely at scaling parameter $\ell$ (dashed lines) are 
integrated out.}
\label{fig:rg}
\end{figure}

Elimination of high energy modes is carried out iteratively, in 
infinitesimal steps, and as a result the energy cutoff $\Lambda$ around the Fermi 
surface shrinks, see Fig. \ref{fig:rg}.  The initial energy cutoff is taken to be
the full band width $\Lambda_0$, and it is reduced via continuous mode elimination 
to $\Lambda = \Lambda_0 ~e^{-\ell}$ where $\ell > 0$. 
For each infinitesimal step $\ell \rightarrow \ell + d\ell$, the fermion degrees of 
freedom are broken down into high and low energy modes as
\begin{eqnarray}
\Psi_{\sigma}(K) = \Theta(\Lambda-|\epsilon_{\bf k}|) ~ \Psi_{\sigma,L}(K) ~+~
          \Theta(|\epsilon_{\bf k}|-\Lambda) ~ \Psi_{\sigma,H}(K) ~,
\end{eqnarray}
where $K \equiv (\omega, {\bf k})$ is the usual 2+1-dimensional frequency-momentum vector.  
The effective action, 
after dropping a constant contribution $\Omega_H$ to the free energy, has the form
\begin{eqnarray}
S_{\Lambda(\ell+d\ell)} = S_{\Lambda(\ell)} + \delta S(\ell).
\end{eqnarray}
At non-zero $\ell > 0$ the effective action 
contains contributions at all orders in the initial interaction strength.  
But because mode elimination is done in infinitesimal 
steps, only terms linear in $d\ell$ contribute to $\delta S(\ell)$.  These terms 
correspond to diagrams with one internal line (either a loop or 
a tree diagram).  RG flow equations for vertices with any number of legs, 
$\Gamma_{2n}(\ell)$, can then be found.  These are functional equations since 
the $\Gamma$'s are functions of momenta and frequencies.

To make progress we must make an approximation.
We carry out the weak-coupling expansion by truncating the RG equations 
at the one-loop level.  Renormalization of the effective interaction 
$U_\ell({\bf k_1}, {\bf k_2}, {\bf k_3}, {\bf k_4})$, corresponding 
to the four-point function ($\Gamma_4$), then occurs at order $U^2$.  
Contributions from the six-point functions $\Gamma_6$ must also be included
at this order.  Higher n-point functions may be neglected as
these only contribute at higher-order in the interaction strength $U$. 
It is important to notice that the RG flow equations generated this way 
are non-local in scaling parameter $\ell$.  The RG equations for couplings 
$U_\ell({\bf k_1}, {\bf k_2}, {\bf k_3}, {\bf k_4})$ at step $\ell$ involve 
the values of couplings at previous steps $\ell_{pp}$ and $\ell_{ph}$ [the 
subscript denotes particle-particle (pp) and particle-hole (ph) channels]: 
\begin{eqnarray}
\ell_{pp} = - \ln \left( \frac{\epsilon_{{\bf k} - {\bf q_{pp}}}}{\Lambda_0} 
\right)\\
\ell_{ph} = - \ln \left( \frac{\epsilon_{{\bf k} + {\bf q_{ph}}}}{\Lambda_0} 
\right)
\end{eqnarray}
with ${\bf q_{pp}} = {\bf k_1} + {\bf k_2}$ and ${\bf q_{ph}} = {\bf k_1} - 
{\bf k_4}$.  At step $\ell$ contributions from six-point functions are
obtained by contracting two of the legs at on-shell momentum 
${\bf k}$.  Of course the six-point functions were generated 
from four-point functions during previous steps.  Momentum ${\bf k_4}$ is 
determined uniquely by momentum conservation to be ${\bf k_4} = {\bf k_1} + 
{\bf k_2} - {\bf k_3}$, so we drop it in the following.

For an initial, bare, four-fermion 
interaction $U_0$ which is independent of spin, following Zanchi and Schulz 
it is possible\cite{Zanchi} to write all the renormalized 
two-particle interactions in terms of 
only one function $U_\ell({\bf k_1}, {\bf k_2}, {\bf k_3})$.  Couplings 
in the charge and spin sectors can then be obtained from this function through 
the relations:
\begin{eqnarray}
U_c = \frac{1}{4} (2-\hat{X}) U, \hskip 1in U_{\sigma} = - \frac{\hat{X}}{4} U.
\end{eqnarray}
where $\hat{X}$ is a permutation operator defined by its action:
$\hat{X} U({\bf k_1}, {\bf k_2}, {\bf k_3}) \equiv 
U({\bf k_2}, {\bf k_1}, {\bf k_3})$.  The charge density (CDW) 
and spin density (AF) couplings are then given by:
\begin{eqnarray}
V_\ell^{CDW}(\theta_1, \theta_2) &=& 4 U_{c \ell}({\bf k_1}, {\bf k_2}, 
{\bf \tilde{k}_1)} 
\nonumber \\
V_\ell^{AF}(\theta_1, \theta_2) &=& 4 U_{\sigma \ell}({\bf k_1}, 
{\bf k_2}, {\bf \tilde{k}_1)}\ .
\end{eqnarray}
Here ${\bf \tilde{k}}_j$ is related to ${\bf k}_j$ by 
${\bf k}_j - {\bf \tilde{k}}_j = {\bf Q}$ where ${\bf Q}$ is a
nesting vector [${\bf Q} = (\pm \pi, \pm \pi)$ for the fully nested 
square lattice].  Also $\theta_j$ is the angle that wavevector ${\bf k}_j$ makes with 
the x-axis.  The forward scattering amplitude is given by 
\begin{equation}
F_\ell(\theta_1, \theta_2) = U_\ell({\bf k_1}, {\bf k_2}, {\bf k_1})
\end{equation}
and only involves two momenta (${\bf k_1}$ and ${\bf k_2}$) because the momentum
transfer during scattering is very small.  Likewise the BCS interaction
\begin{equation}
V_\ell^{BCS}(\theta_1, \theta_2) = U_\ell({\bf k_1}, -{\bf k_1}, {\bf k_2}) 
\label{eq:vbcs}
\end{equation}
also is described by just two momenta as it represents the scattering of a Cooper
pair of electrons of opposing momenta ${\bf k_1}$ and ${\bf -k_1}$ 
into a pair of electrons of opposing momenta ${\bf k_2}$ and ${\bf -k_2}$.
 
\begin{figure}[!ht]
\includegraphics[width=2in]{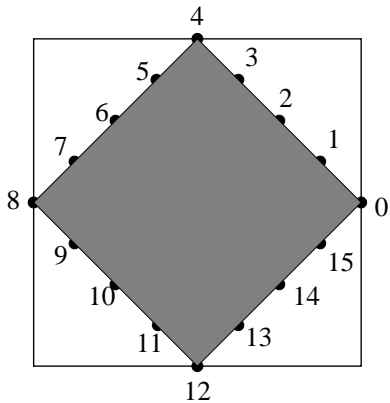}
\caption{Discretization of the Fermi surface into $M = 16$ patches.  Each patch 
corresponds to an angular section of $2\pi/M$.  The special case of the 
perfectly-nested Fermi surface corresponding to the nearest-neighbor tight-binding
model on a square lattice and at half-filling is shown.}
\label{fig:discrete}
\end{figure}

In order to integrate the flow equations forward in the scaling parameter
$\ell$ we first discretize the Fermi surface, dividing it up into patches 
as depicted in Fig. \ref{fig:discrete}.  Replacing the continuous surface
with discrete patches should be adequate for the imperfectly nested Fermi surfaces
we focus on here\cite{Doucot}.  After discretization of the Fermi surface, 
the angles $\theta_j = 2 \pi j / M$ where $j = 0, \ldots, M-1$.  
Interactions $U_\ell(i_1,i_2,i_3)$ are thus labeled by three discrete patch 
indices.  A further approximation is implied by this 
procedure, as the dependence of the effective interaction on the radial 
component of momentum is neglected and the shape of the Fermi surface is 
not renormalized.  The justification is the following: though the shape of the 
Fermi surface change at the one-loop level, the feedback of this change on the 
one-loop RG flows for the couplings $U_\ell(i_1,i_2,i_3)$ constitutes a 
higher-order effect.  The dependence of $U$ on the radial components of the 
three momenta is irrelevant\cite{Zanchi2,Shankar}.  This is similar to the 
one-dimensional case, where the marginal interactions are labeled according to 
the indices $i = L, R$ (left or right moving) of the electrons that are 
interacting.  There is strong dependence on the direction of ${\bf k}$, 
but the dependence on the absolute value $|{\bf k}|$ of the momentum is irrelevant.  
Therefore the interactions may be parameterized simply
by their projection onto the two Fermi points.  
In two dimensions the interactions are likewise parameterized by the 
patch indices.  

In this work we only study flows at zero temperature.  The integral over 
Matsubara frequencies, which arises in the one-loop diagram, 
can be performed analytically as the dependence of the couplings 
on the frequency $\omega$ is irrelevant\cite{Zanchi2}.  We set the initial bare 
coupling to be $|U_0| = 1$ and, unless otherwise stated, also set $t_1 + t_2 = 1$. 
The full band width is $\Lambda_0 = 8 t_1 + 4 t_2$. 
We usually divide the Fermi surface into $M = 16$ patches.  For the special case of
the isotropic triangular lattice we instead use a finer mesh of patches, $M = 24$, 
to permit an examination of higher-wave channels.  
Our algorithm makes no assumptions about the symmetries of the 
Fermi surface; this means that we must follow the flow of all $M^3$ 
couplings $U(i_1,i_2,i_3)$.  We do impose the requirement that the three indices 
are such that all four particles lie on the Fermi surface.  
The RG flow for these couplings are then described by 
coupled non-local integral-differential equations.  These equations are 
numerically integrated forward in the scaling parameter $\ell$.  
The increment in the scaling parameter is set to be $d\ell = 0.1$ 
for the results shown here.  Calculations using smaller values of 
$d\ell$ yield nearly the same results.

An equivalent version of RG method for two-dimensional 
interacting fermions has been developed by 
Salmhofer\cite{Salmhofer1,Salmhofer2}.  In this formulation, the RG flow 
equations are {\it local} in the scaling parameter $\ell$, but
this gain comes at the cost of expanding the effective action in 
{\it Wick-ordered} monomials, resulting in RG flow equations with one 
extra integration over momentum.  This formulation has been used to study the 
two-dimensional Hubbard model on a square lattice with nearest-neighbor and 
next-nearest-neighbor hopping amplitudes\cite{Halboth1,Halboth2,Honerkamp}.

Given a set of RG flows, we must then interpret the various ordering tendencies.
One way to do this is by calculating susceptibilities towards order, as carried out
for instance in Refs. \onlinecite{Halboth1,Halboth2}.  Another approach is 
to bosonize the fermion degrees of freedom, and then determine the ground
state of the bosonized effective Hamiltonian semiclassically by replacing 
each boson field with
a c-number expectation value.  The latter method was adopted by Lin, Balents,
and Fisher in their treatment of the two-leg ladder system\cite{LBF98}.  Klein
ordering factors must be treated carefully\cite{John} and the resulting
weak-coupling RG / bosonization prediction was shown to agree well with the
results of essentially exact DMRG calculations\cite{MFS}.  We leave the 
extension of such an analysis to the full two-dimensional problem\cite{HKM} 
for future work, and make the observation here that in most instances 
it suffices to simply follow, during the course of the RG flow,
the most rapidly diverging interaction channel.  For instance, 
the effective BCS interaction $V_\ell^{BCS}(i_1, i_2)$, as 
defined by Eq. \ref{eq:vbcs}, is a symmetric $M \times M$ matrix in the patch indices.
The various BCS channels are obtained upon diagonalizing the matrix.  The eigenvector $\phi$ 
with the largest attractive eigenvalue then represents the dominant BCS channel.
We also calculate the eigenvectors and eigenvalues of the 
effective spin coupling $V_\ell^{AF}(i_1, i_2)$ and charge density wave coupling 
$V_\ell^{CDW}(i_1, i_2)$ to determine the dominant AF and CDW channels.  
In the following we plot largest eigenvalues of the interaction matrices 
as a function of the scaling parameter, as well as the dominant eigenvectors as
a function of the patch index, to gain insight into the ordering tendencies.  
As shown in the next section this way of intepreting the RG flows 
yields the correct physics in the limiting cases of one-dimensional decoupled
chains as well as the completely nested square lattice.

\section{Results}
\label{sec:results}

We now turn to the results of our RG calculation.  We first check the method in 
the special limiting cases of decoupled chains and the pure square lattice.  As we 
reproduce known results in these limits, we then turn to the more general problem
of the anisotropic triangular lattice.

\subsection{Decoupled chains ($t_1 = 0$)}
\label{subsec:chains}

As a first check, we apply the weak-coupling analysis to the case 
$t_1 = 0$ and $t_2 = 1$ which corresponds to completely decoupled 
chains.  At half-filling, particle-hole symmetry requires $\mu = 0$
and the nesting wavevector is $Q = \pi$.  As quantum fluctuations always suffice 
to prevent continuous symmetries from breaking in one spatial dimension, 
antiferromagnetic and superconducting order are not possible.  Instead the possible
phases are classified in terms of whether or not charge and/or spin excitations
are gapped.  Furthermore the three types of marginal interactions are often 
denoted (see Ref. \onlinecite{Affleck}) spin current ($\lambda_s$), 
charge current ($\lambda_c$), and Umklapp ($\lambda_u$), where the latter two carry only
charge and no spin. In terms of our notation we may identify 
\begin{eqnarray}
\lambda_c &=& 4 U_c(R, L, R), \nonumber \\
\lambda_s &=& 4 U_{\sigma}(R, L, R), \\
\lambda_u &=& U(R, R, L)\ . \nonumber
\end{eqnarray}

\begin{figure}[!ht]
\includegraphics[width=2.5in,clip]{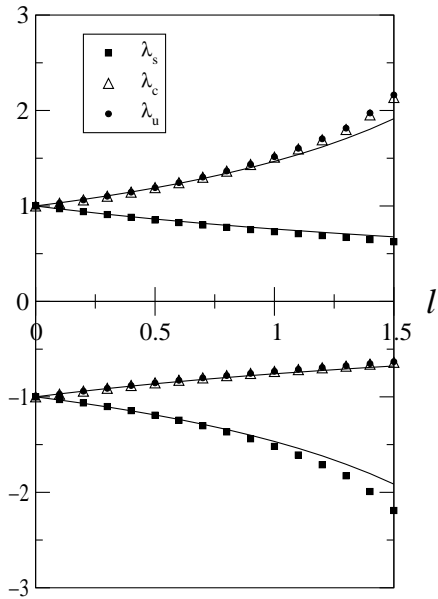}
\caption{RG flows of the spin current ($\lambda_s$), charge current 
($\lambda_c$) and 
Umklapp ($\lambda_u$) interactions, in the limit of completely 
decoupled chains ($t_1 = 0$).  
For repulsive (attractive) bare interaction $U_0 > 0$ ($U_0 < 0$), a charge (spin) 
gap develops and there are gapless spin (charge) excitations.  Solid lines are the
analytical solution of the one-loop RG equations for the one-dimensional Hubbard model
at half-filling.}
\label{fig:chains}
\end{figure}

As shown in Fig. \ref{fig:chains}, 
for repulsive initial interaction ($U_0 > 0$), the spin 
couplings decrease towards zero, whereas the Umklapp and charge
couplings diverge in the low-energy limit.  This is as expected from the
exact Bethe ansatz solution\cite{Lieb} since the system has gapless 
spin excitations while the charge sector is gapped.  On the other hand, for
attractive initial interaction ($U_0 < 0$) the spin couplings diverge while 
Umklapp and charge couplings tend towards zero.  In this case there 
is a gap in the spin sector and gapless charge excitations.   
The solid lines in Fig. \ref{fig:chains} 
correspond to a direct analytical solution of the simple one-loop RG equations for the 
one-dimensional Hubbard model at half-filling:
\begin{eqnarray}
{d\lambda_s \over d\ell} &=& - {1\over \pi} \lambda_s^2, \nonumber \\
{d\lambda_c \over d\ell} &=& {1\over \pi} \lambda_u^2, \nonumber \\
{d\lambda_u \over d\ell} &=& {1\over \pi} \lambda_c \lambda_u\ . 
\end{eqnarray}
Our numerical solution of the 
Zanchi-Schulz RG equations agrees quantitatively with the standard one-loop results.  
An exact fit is not expected, because the the Zanchi-Schulz equations also include 
the renormalization of the charge and spin speeds as well as 
sub-leading non-logarithmic corrections.

\subsection{Square lattice ($t_2 = 0$)}
\label{subsec:square}

\begin{figure}[!ht]
\includegraphics[width=3in]{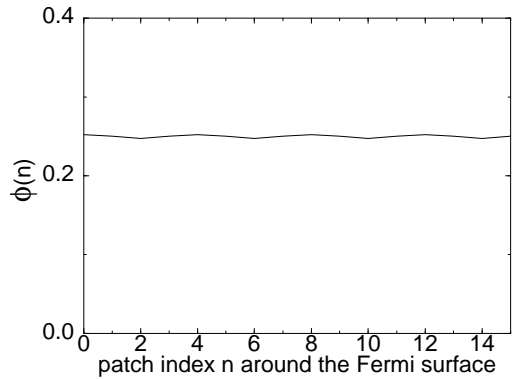}
\caption{Pairing symmetry of the dominant BCS channel for the case of 
attractive initial interaction ($U_0 < 0$) on a square lattice 
($t_1 = 1$ and $t_2 = 0$) and at half-filling ($\mu = 0$), as revealed by
plotting the eigenvector of the BCS matrix with the largest attractive 
eigenvalue.  The scaling parameter $\ell = 2.5$.   
The Fermi surface is divided in $M = 16$ patches.  As the patch index goes from $n = 0$ 
to $n = 15$ around the Fermi surface, the angle $\theta$ goes from $0$ to $2 \pi$.}
\label{fig:swave}
\end{figure}

\begin{figure}[!ht]
\includegraphics[width=3in,clip]{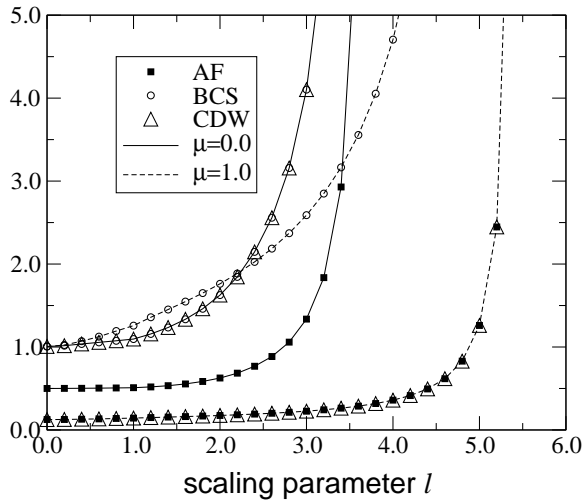}
\caption{RG flows of the dominant AF, BCS and CDW couplings for the square lattice with 
an attractive Hubbard interaction ($U_0 < 0$).  Plotted are the largest eigenvalues of
the corresponding interaction matrices.  Solid lines correspond 
to the half-filled case ($\mu = 0.0$).  Nesting occurs at half-filling and 
$V^{CDW}$ diverges as strongly as $V^{BCS}$.  Dashed lines correspond to 
$\mu = 1.0$.  In this case, nesting is destroyed causing $V^{AF}$ and 
$V^{CDW}$ to increase at a much smaller rate.}
\label{fig:mu01}
\end{figure}

\begin{figure}[!ht]
\includegraphics[width=3in,clip]{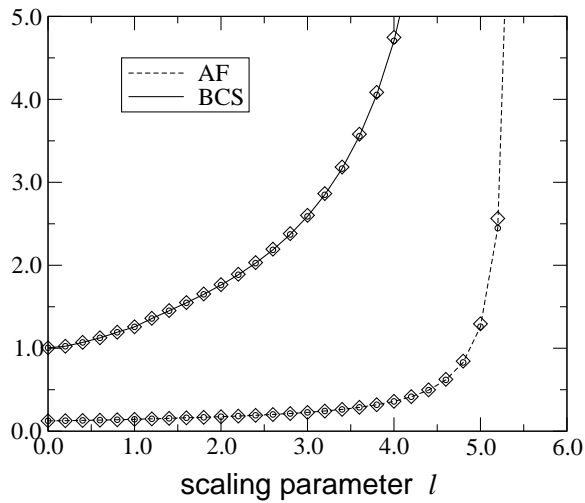}
\caption{Comparison of results obtained from the approach of Zanchi and Schulz
(circles) with those obtained when only logarithmic contributions are 
included (diamonds), for the case of an attractive Hubbard 
interaction ($U_0 < 0$) and $\mu = 1$. There is no significant 
discrepancy between the results.  Flow in the dominant 
BCS coupling is shown by the solid line; the AF coupling is indicated
by the dashed line.}
\label{fig:negu}
\end{figure}

Attractive interactions $U_0 < 0$ induce strong BCS instabilities in the case 
$t_2 = 0$ of a pure square lattice, in accord with 
expectations.  The eigenvector of the dominant attractive BCS channel   
is plotted in Fig. \ref{fig:swave} for the case of half-filling, $\mu = 0$.  
As expected, the BCS pairing is in the $s$-wave channel when $U_0 < 0$.  From the 
outset at $\ell = 0$ the BCS sector dominates all other channels.  As
the RG flows progress, the BCS channel diverges and thus remains the dominant 
coupling.  The same qualitative behavior persists as the system is doped away
from half-filling.  
Fig. \ref{fig:mu01} depicts the RG flows in the dominant AF, BCS and CDW channels 
both for the half-filled case $\mu = 0$ and away from half-filling ($\mu = 1$).  
Perfect nesting at half-filling drives $V^{CDW}$ to diverge 
as strongly as $V^{BCS}$.  Away from half-filling, there is no perfect nesting 
so both $V^{AF}$ and $V^{CDW}$ grow at a much smaller rate.  In Fig. 
\ref{fig:negu}, we compare our results to RG flows which include only 
the leading logarithmic contributions.  As the BCS coupling is not driven here by 
the AF fluctuations, there is little coupling between the two channels.  Therefore, 
one-loop RG flows which include only the leading logarithmic contributions (diamonds) 
do not deviate significantly from the more accurate approach of Zanchi and Schulz 
(circles) which includes all the subleading non-logarithmic terms 
generated at one-loop.

\begin{figure}[!ht]
\includegraphics[width=3in]{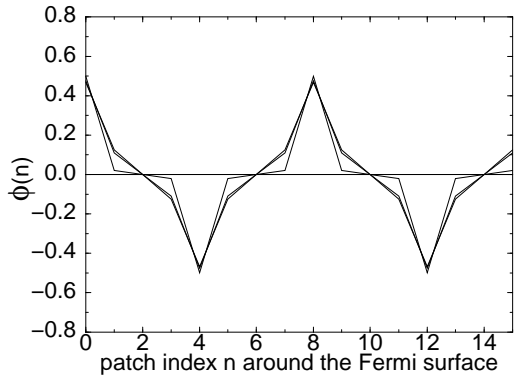}
\caption{Pairing symmetry of the dominant BCS channel for the case of 
repulsive initial interaction ($U_0 > 0$) on the square lattice 
($t_1 = 1$ and $t_2 = 0$) at half-filling ($\mu = 0$).  The different 
curves correspond to different values of the scaling parameter 
($\ell = 0.5, 1$ and $3$).  The 
Fermi surface is divided in $M = 16$ patches.  As the patch index increases from 
$n = 0$ to $n = 15$ around the Fermi surface, the angle $\theta$ increases from 
$0$ to $2 \pi$.} 
\label{fig:dwave}
\end{figure}

The opposite limit of a square lattice with repulsive initial interaction 
$U_0 > 0$ has been extensively studied\cite{Zanchi,Halboth1}.  At 
half-filling, the 
Fermi surface is perfectly nested and a strong SDW instability develops.  
Fig. \ref{fig:dwave} shows, however, that the largest BCS channel,
though sub-leading in comparison to the SDW channel, 
already exhibits $d_{x^2-y^2}$ pairing symmetry. 

\begin{figure}[!ht]
\includegraphics[width=3in,clip]{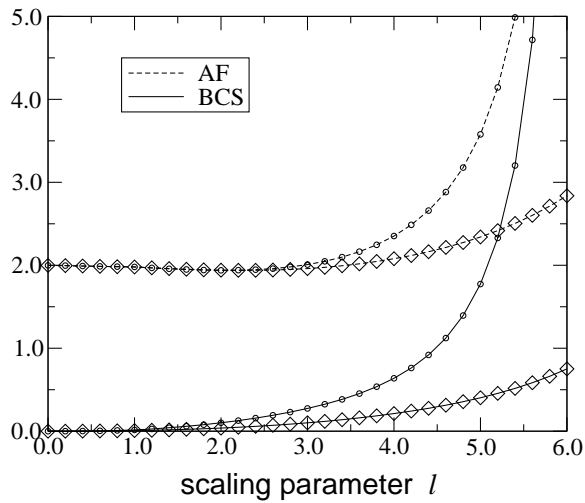}
\caption{Comparison of results obtained at one-loop level for the case of a 
repulsive Hubbard interaction and $\mu = 10^{-4}$.  RG flows which
include the effects of formally irrelevant terms (circles) are compared with 
RG flows for which only logarithmic contributions are retained (diamonds).   
Flow in the dominant BCS channel is depicted as a solid line and 
dominant AF sector is indicated by a dashed line.}
\label{fig:posu}
\end{figure}

For the case of a repulsive Hubbard interaction, $U_0 > 0$, we find
in contrast to the attractive situation that the formally irrelevant
terms play an important role.  As the initial BCS couplings are all
repulsive, Cooper pairing can only happen via coupling to the AF 
channels or via the non-logarithmic corrections to scaling coming from the 
formally 
irrelevant terms in the six-point functions.  We again compare RG flows which 
include only the leading logarithmic corrections (diamonds) 
against those in which all subdominant contributions at one-loop order are included
(circles) in Fig. \ref{fig:posu} for the case $\mu = 10^{-4}$, that is, 
slightly away from half-filling.  Though qualitatively similar, there is 
considerable quantitative difference.  At this small doping, AF tendencies dominate
in both cases.  Upon further increasing the chemical potential, as mentioned 
above there is a crossover into the $d_{x^2 - y^2}$ BCS regime.  The 
crossover occurs much sooner
when the subleading terms are included.  Finally, at large doping 
the nesting of the Fermi surface is completely eliminated, and 
neither the AF nor the BCS channels show any strong divergences.
 
\subsection{Square Lattice With Next-Nearest-Neighbor Hopping ($t^\prime \neq 0$)}
 
\begin{figure}[!ht] 
\includegraphics[width=1.5in]{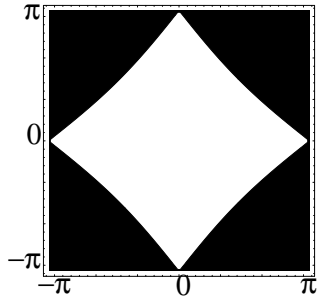}
\caption{Fermi surface of the square lattice with nearest-neighbor hopping 
$t_1 = - 1$ and next-nearest-neighbor hopping $t^\prime = 0.05$.  The 
chemical potential $\mu = 4 t^\prime$.}
\label{fig:vhfs}
\end{figure}

Next we turn to the square lattice with added next-nearest-neighbor hopping 
amplitude $t^\prime$ along each of the two diagonal directions.  
Weak-coupling RG studies of this Hubbard model 
have been carried out previously by a number of 
groups\cite{Halboth1,Halboth2,Honerkamp} using the formulation of 
Refs. \onlinecite{Salmhofer1,Salmhofer2}. We have checked our calculation 
against these published results and find good agreement.  
The dispersion relation in this case is given by
\begin{eqnarray}
\epsilon_{\bf k} = - 2 t_1 (\cos k_x + \cos k_y) - 4 t^\prime \cos k_x \cos k_y
\end{eqnarray}
and Fig. \ref{fig:vhfs} show the Fermi surface for the case $t_1 = - 1$, $t^\prime = 0.05$,
and $\mu = 4 t^\prime$.  The Fermi surface is centered around the $\Gamma$ point
$(\pm \pi, \pm \pi)$ and van Hove singularities lie at the Fermi energy\cite{Gonzalez}. 

\begin{figure}
\includegraphics[width=3in]{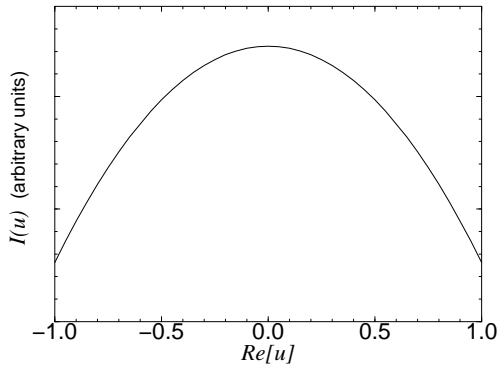}
\caption{Dependence of the integral $I(u)$ on the real part of $u$. The maximum 
of $I(u)$ occurs when Re$\{u\} = 0$, that is, when $u$ is purely imaginary.}
\label{fig:i}
\end{figure}

This Fermi surface exhibits the $d_{x^2-y^2}$ superconducting instability.  
It is interesting to go a bit further and address the question of whether or not
there is spontaneous time-reversal ($\hat{T}$) symmetry breaking with 
the appearance of an additional imaginary $i d_{xy}$ component to the superconducting
order parameter.  To answer this question we study the implications of RG flows which 
yield comparable attraction in two channels: one term with $d_{x^2-y^2}$ symmetry and a second
with $d_{xy}$ symmetry.  A simple calculation of energetics then suffices to show that 
the two order parameters will phase-align as $d_{x^2-y^2} + i d_{xy}$. 
The standard BCS equation yields a condensation energy of
\begin{eqnarray}
\Delta E = E_{SC} - E_N = 2 \sum_{{\bf k} > {\bf k_F}} 
\left[ \epsilon_{\bf k} - \frac{2 \epsilon_{\bf k}^2 + 
\Delta_{\bf k}^2}{2 \sqrt{\epsilon_{\bf k}^2 + 
\Delta_{\bf k}^2}} \right]\ . 
\label{eq:bcs}
\end{eqnarray}
For couplings $V^{BCS}$ with comparable $d_{x^2-y^2}$ and $d_{xy}$ components, 
the ansatz to maximize the condensation energy should be chosen to be
$\Delta_{\bf k} = \Delta_{d_{x^2-y^2}}({\bf k}) + 
u~ \Delta_{d_{xy}}({\bf k})$, with $u$ encoding information about the 
relative phase of the two components.  Substituting this ansatz into 
Eq. \ref{eq:bcs}, we may then determine the phase that maximizes the condensation 
energy $\Delta E$.  Fig. \ref{fig:i} shows the dependence of the sum 
\begin{eqnarray}
I(u) = \sum_{{\bf k} > {\bf k_F}} {2 \epsilon_{\bf k}^2 + \Delta_{\bf k}^2
\over 
\sqrt{\epsilon_{\bf k}^2 + \Delta_{\bf k}^2}}
\end{eqnarray} on the real part of $u$.  This term is maximized when $u$ 
is purely imaginary (Re$\{u\} = 0$), hence the $\hat{T}$-breaking 
pairing symmetry $d_{x^2-y^2} + i d_{xy}$ is the energetically 
favored.  Physically this is reasonable, as this choice of the phase 
guarantees that a gap forms everywhere along the Fermi surface, lowering
the ground-state energy.  

\begin{figure}[!ht]
\centerline{\includegraphics[width=3in]{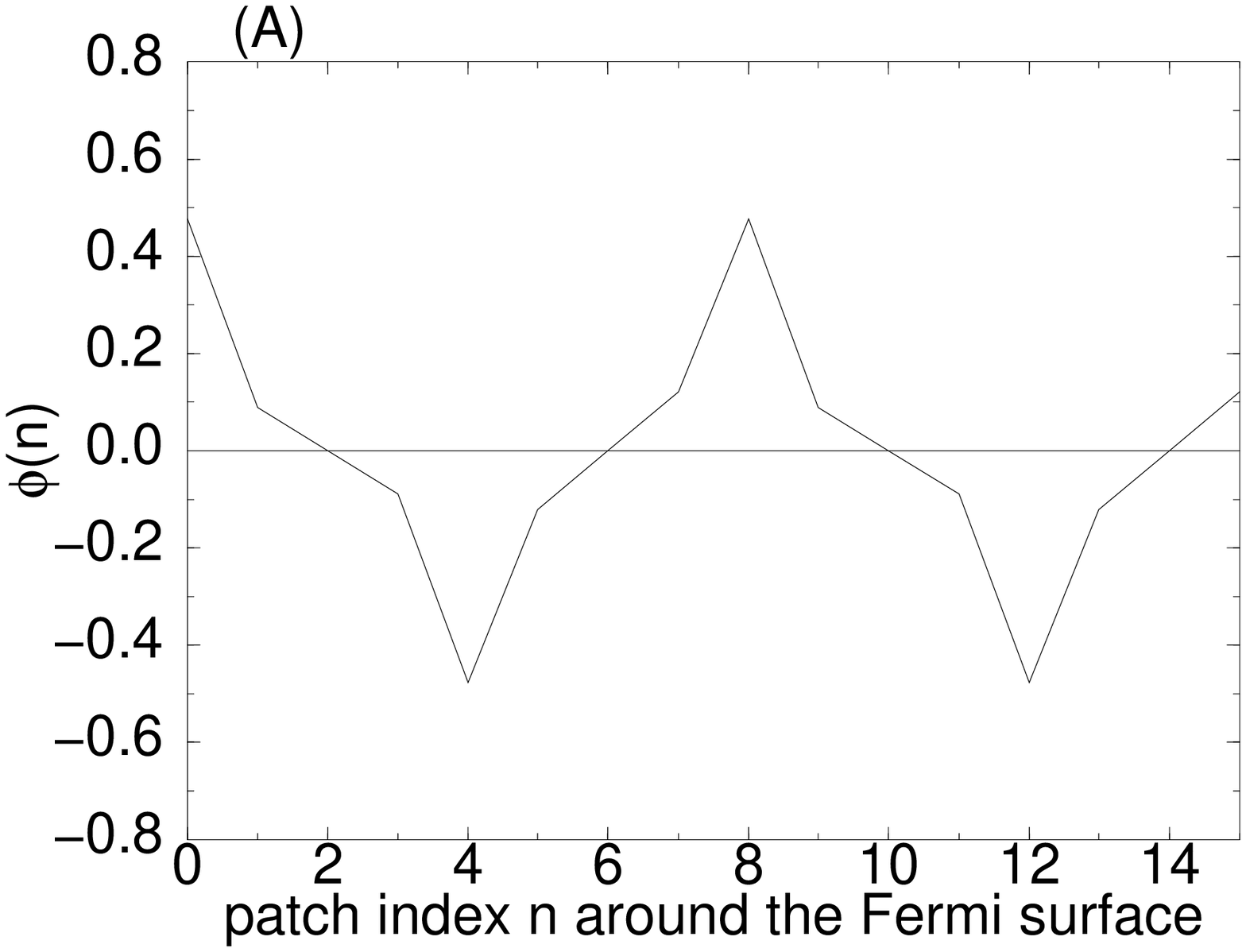}
\includegraphics[width=3in]{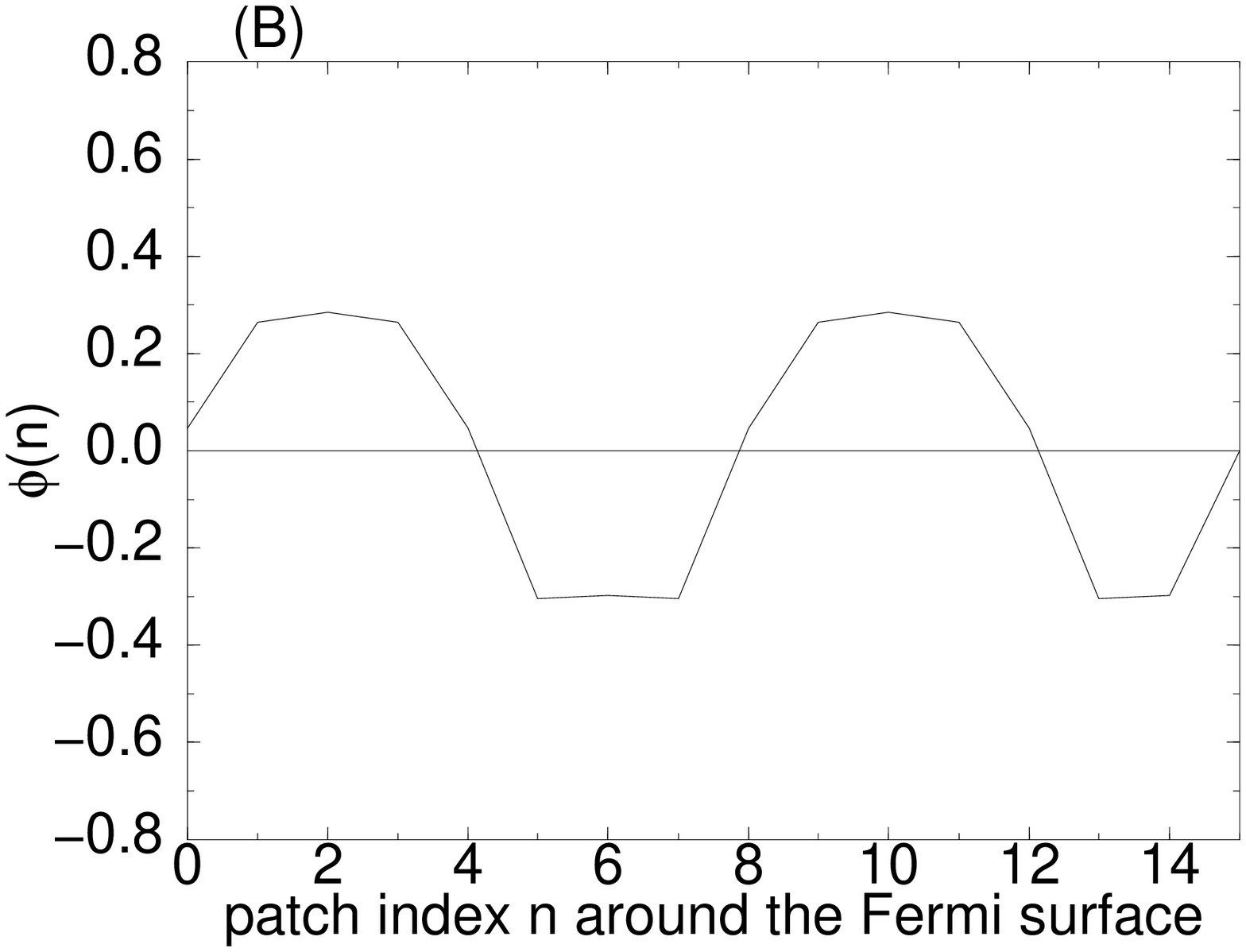}}
\caption{Dominant BCS channels for the square lattice with $t_1 = - 1$, 
$t^\prime = 0.05$ and $\mu = 4 t^\prime$, at scaling parameter $\ell = 5$.  
The $d_{x^2-y^2}$ channel (A) is attractive 
while the $d_{xy}$ channel (B) is repulsive.  The absolute values of the 
eigenvalues are plotted in Fig. \ref{fig:vhd} as functions of the scaling 
parameter $\ell$.}
\label{fig:vhbcs}
\end{figure}

\begin{figure}[!ht] 
\includegraphics[width=3in]{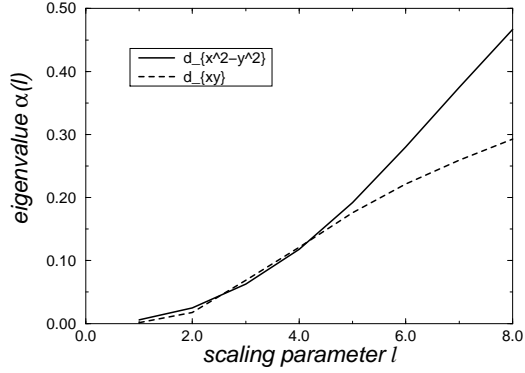}
\caption{Flow of the eigenvalues for the $d_{x^2-y^2}$ and $d_{xy}$ channels 
shown in Fig. \ref{fig:vhbcs}.  The magnitude of the eigenvalues is plotted.}
\label{fig:vhd}
\end{figure}

Returning to the square lattice, we find that upon 
integrating the RG equations for the case $t_1 = - 1$, $t^\prime = 0.05$ and 
$\mu = 4 t^\prime$, the dominant attractive BCS channel has 
$d_{x^2-y^2}$ symmetry as expected; see Fig. \ref{fig:vhbcs}(A).  A channel 
with $d_{xy}$ symmetry also appears but it is repulsive in sign; see Fig. \ref{fig:vhbcs}(B).   
The strengths of each channel are plotted in Fig. \ref{fig:vhd} as a function 
of the scaling parameter $\ell$.  Since the $d_{xy}$ channel is repulsive, no $d_{xy}$
order will arise, and this may be taken as evidence against the formation
of spontaneous time-reversal symmetry breaking of the $d_{x^2-y^2} + i d_{xy}$ type.

\subsection{Towards the Triangular Lattice ($t_2 \neq 0$, $t_2 < t_1$)}
\label{subsec:away}

\begin{figure}[!ht]
\includegraphics[width=3in,clip]{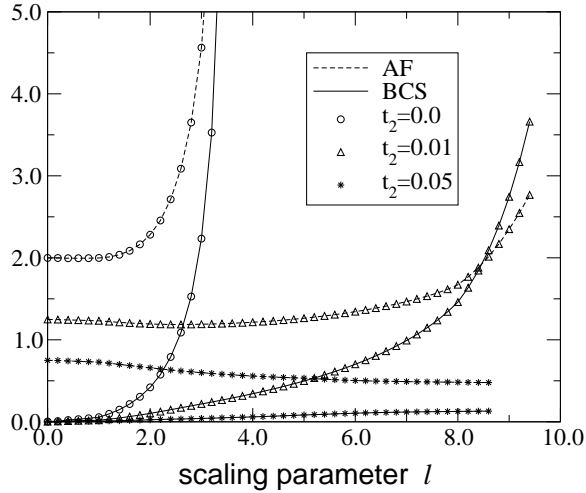}
\caption{At half-filling, spin-wave instability occurs for $t_2 = 0$, but 
as $t_2$ increases, the BCS instability wins over.  This is due to 
imperfect nesting.  As $t_2$ is increased further, the nesting is destroyed 
and both divergences are suppressed.  The hopping $t_1$ is chosen such that 
$t_1 + t_2 = 1$.}
\label{fig:afxsc_t2}
\end{figure}

Introducing non-zero $t_2$ along just one of the two diagonals,
as shown in Fig. \ref{fig:tri},  
offers a {\it different} way of breaking-up perfect
nesting and enhancing BCS instabilities, even at half-filling\cite{Tsai}.  
For sufficiently large $t_2$ there is a crossover to a regime where the BCS 
processes eventually dominate, signaling a superconducting instability.  
Furthermore, because the Fermi surface is imperfectly nested, the growth of
both BCS and AF couplings weakens.  Further increasing $t_2$ eventually 
destroys nesting of the Fermi surface altogether and both types of 
divergences are suppressed.  
Three cases illustrating the crossover are shown in Fig. \ref{fig:afxsc_t2}. 

\begin{figure}[!ht]
\centerline{
\includegraphics[width=3in]{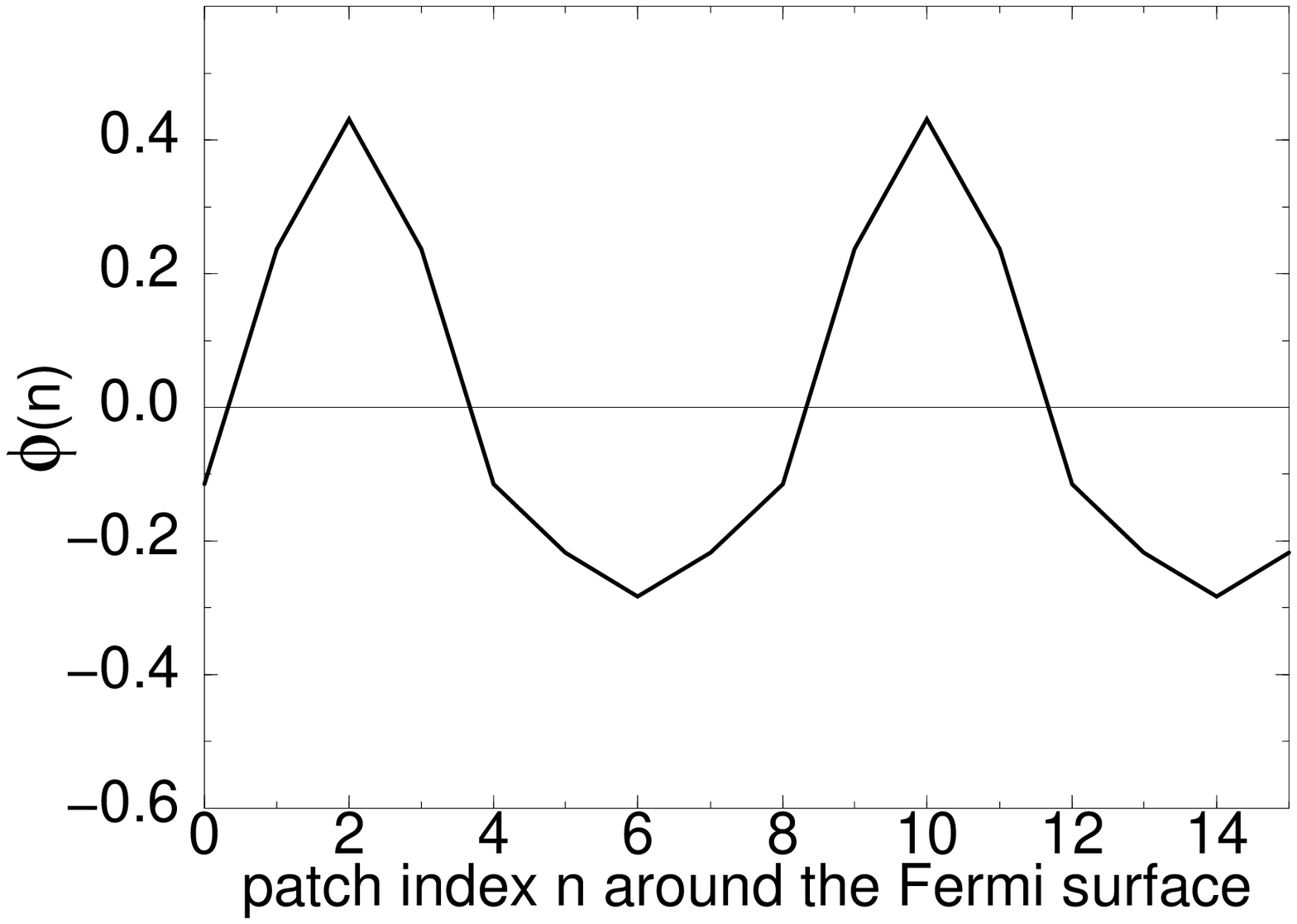}
\includegraphics[width=3in]{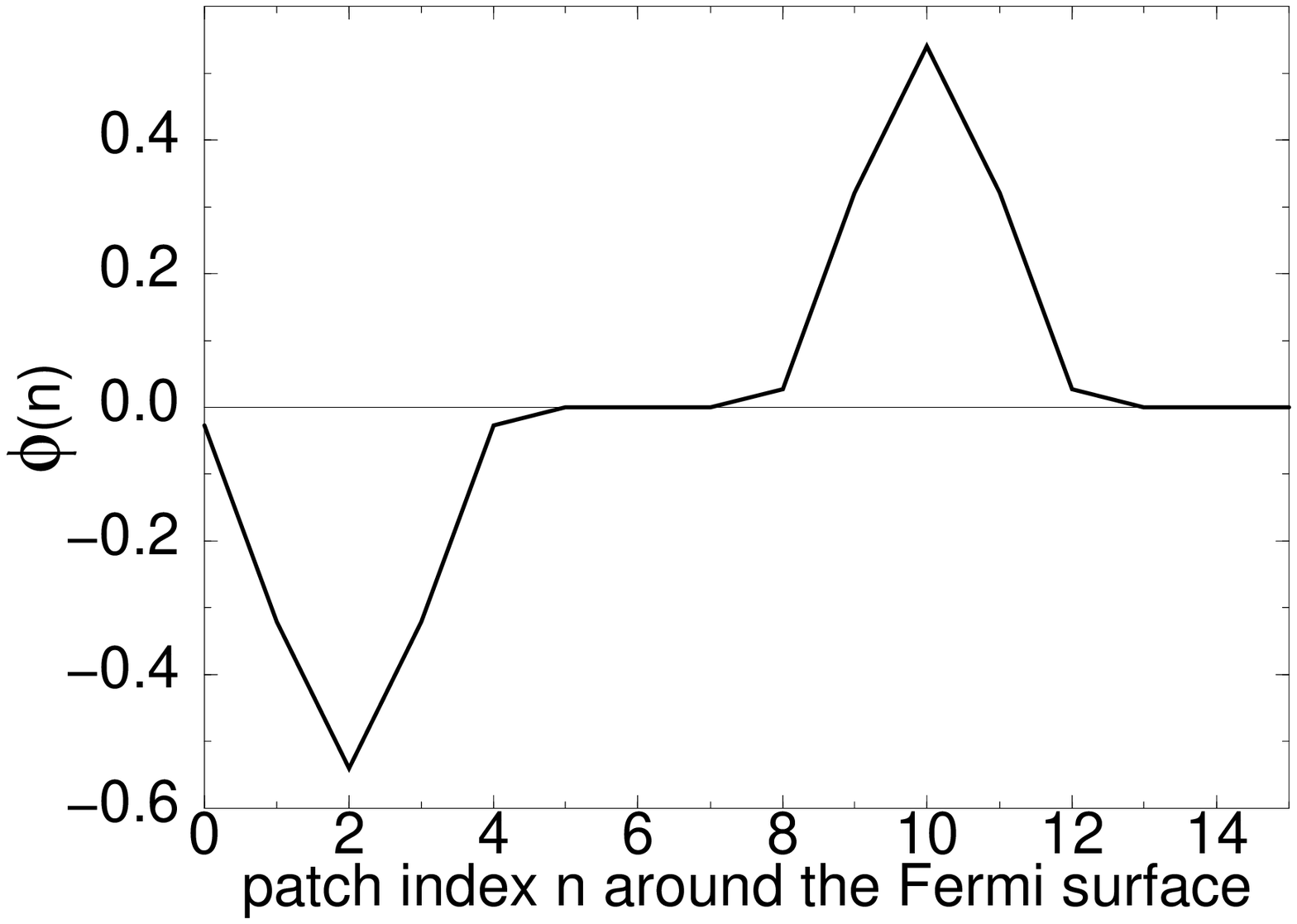}}
\caption{Dominant BCS channels for $t_1 = 0.9$ and $t_2 = 0.1$, at 
half-filling and at scaling parameter $\ell = 5$.  These channels are both
repulsive (with eigenvalues of 0.19525 and 0.162 respectively).  
The attractive $d_{x^2-y^2}$ BCS channel has a small eigenvalue of -0.268.}
\label{fig:repulsive}
\end{figure}

As further increases in the diagonal hopping $t_2$ suppress the $d_{x^2-y^2}$ BCS 
channel, this channel 
diminishes relative to other subdominant BCS channel with different symmetries, 
for example d$_{xy}$- or p-wave, as shown in Fig. \ref{fig:repulsive}.  
These other channels, however, are all repulsive and hence do not 
lead to BCS instabilities by themselves.

We note that the Hubbard model on the anisotropic triangular lattice has 
also been studied using the the random-phase approximation\cite{Vojta} and the
fluctuation-exchange (FLEX) approximation\cite{Kondo,Kino,Schmalian}.  The $d$-wave 
superconducting instability was found to be dominant for a large range of values of 
$t_2/(t_1 + t_2)$ interpolating between the square lattice and the isotropic 
triangular lattice. 

\subsection{Isotropic Triangular Lattice At Half-Filling}
\label{subsec:triangular}

\begin{figure}[!ht]
\centerline{
\includegraphics[width=3in]{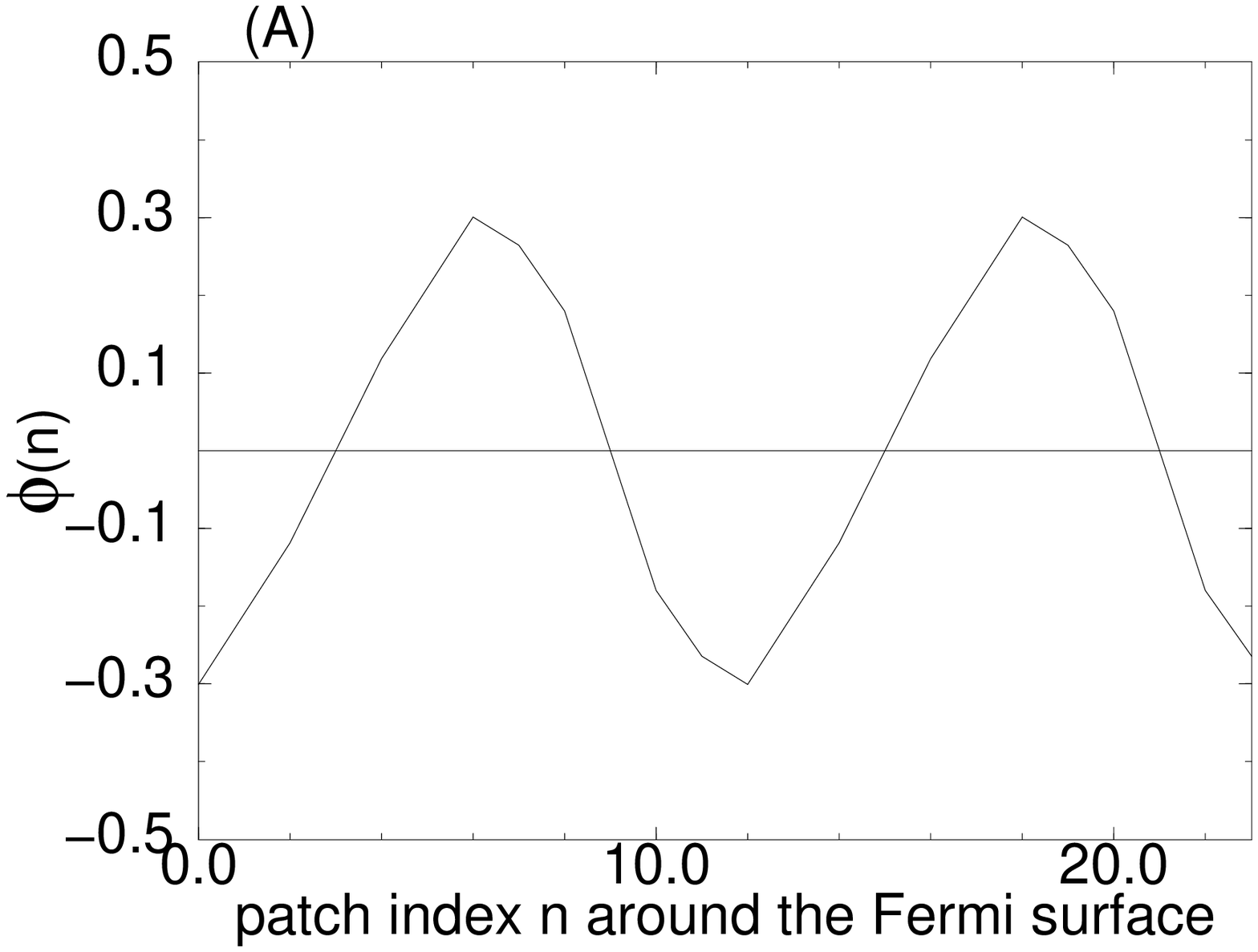}
\includegraphics[width=3in]{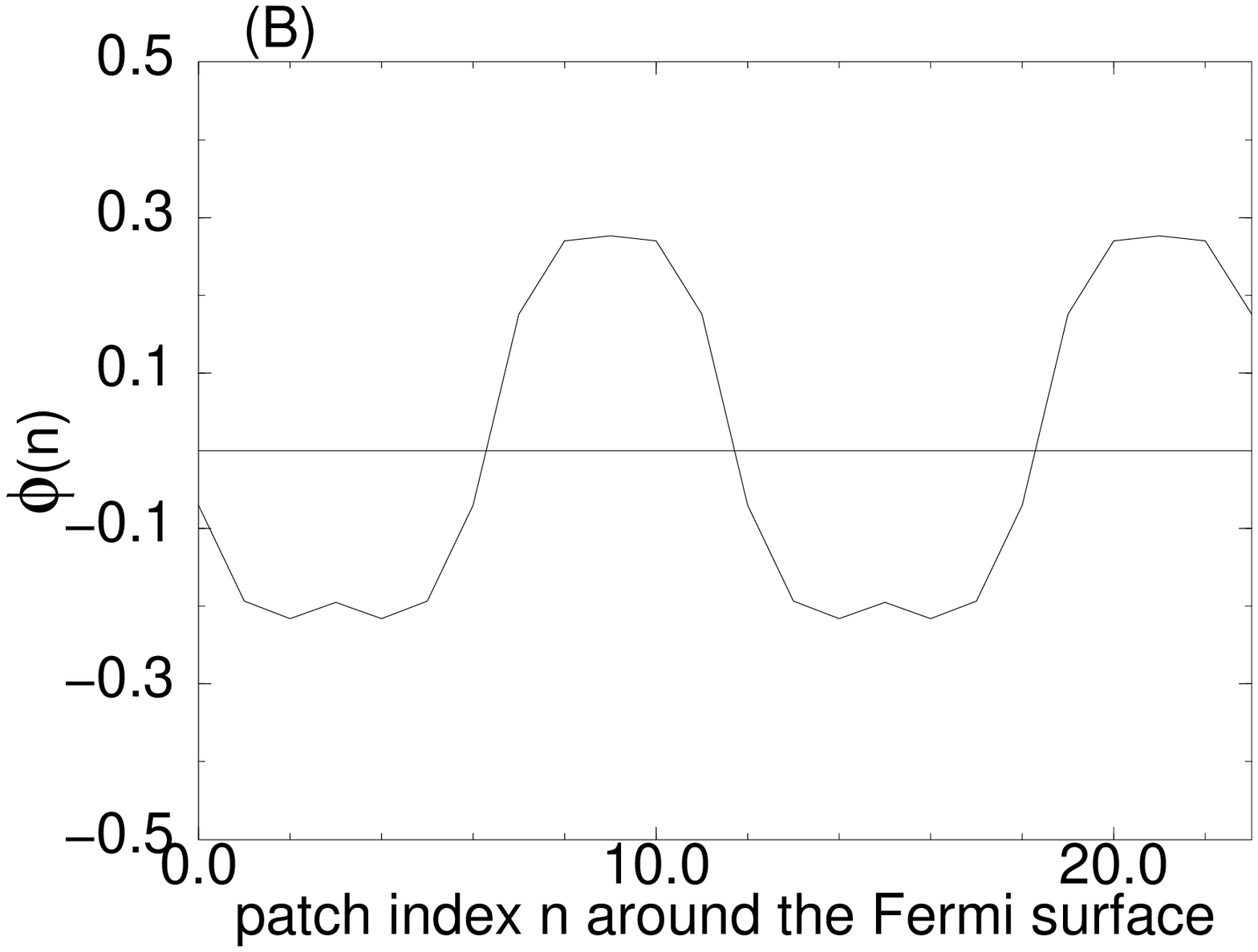}}
\centerline{
\includegraphics[width=3in]{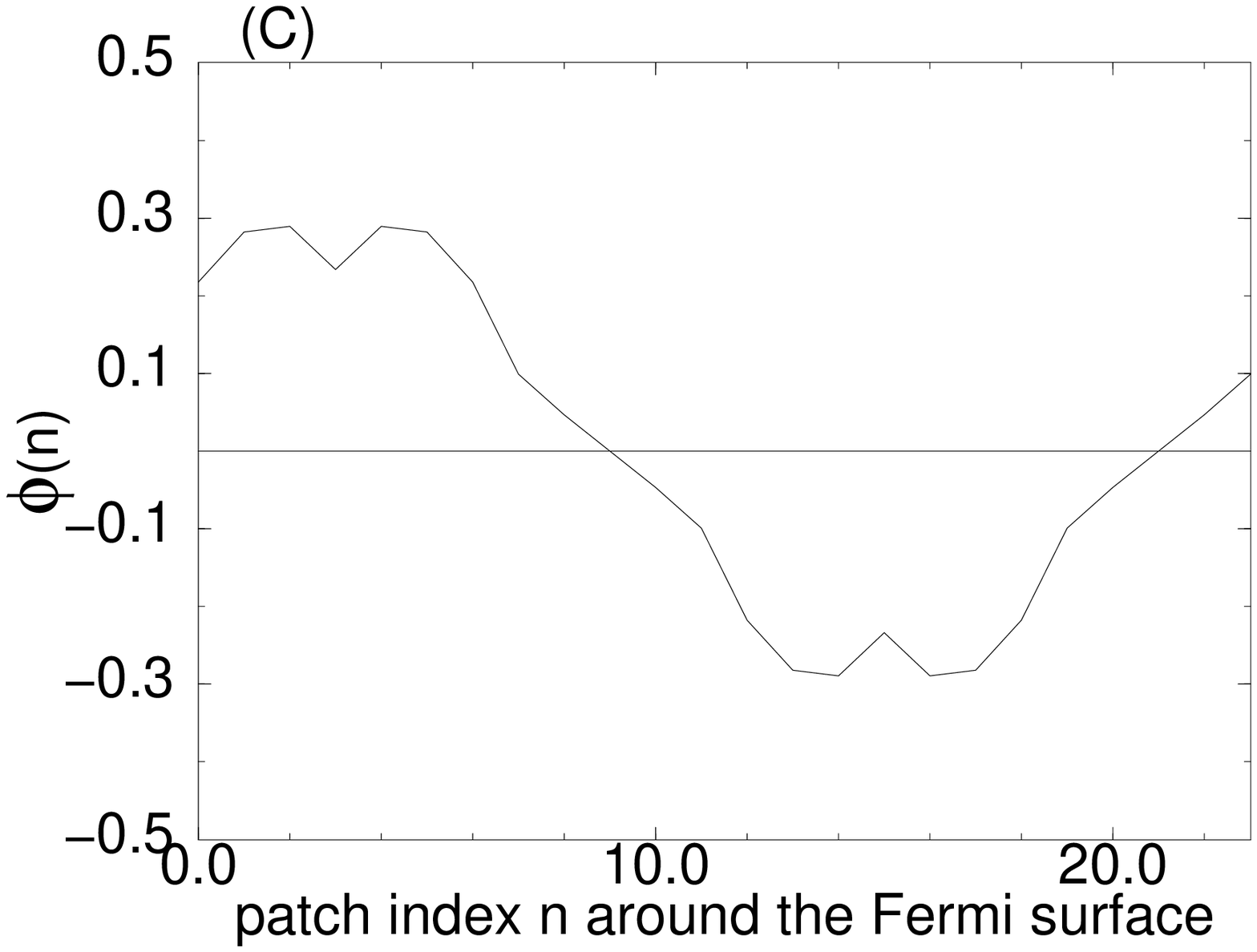}
\includegraphics[width=3in]{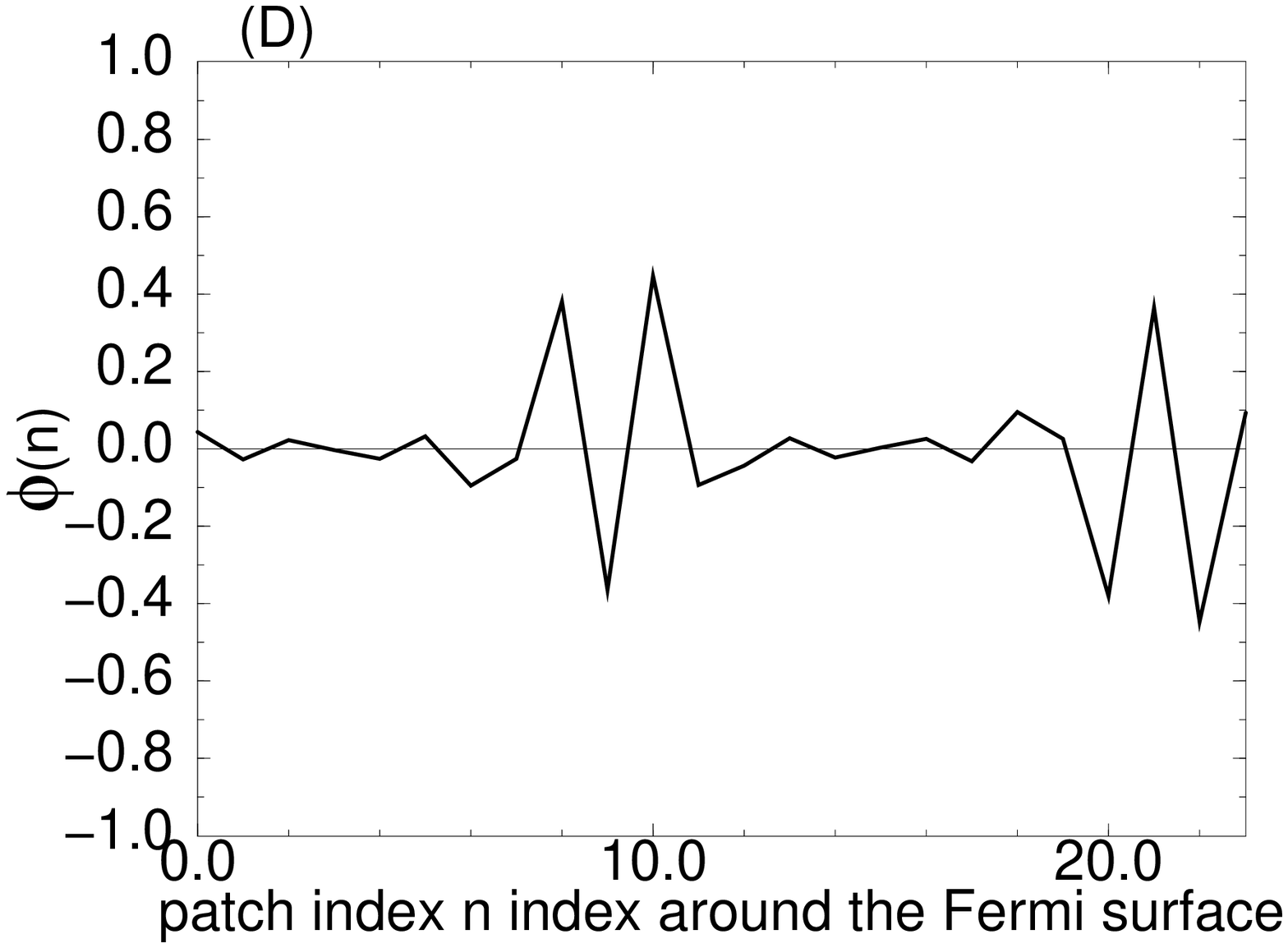}}
\caption{BCS channels for the isotropic triangular lattice 
$t_1 = t_2 = 0.5$, at scaling parameter $\ell = 8.4$.  Channels with $d_{x^2-y^2}$ (A), $d_{xy}$ (B) and $p$ 
(C) symmetries all appear as repulsive channels (with eigenvalues 0.0378, 
0.035 and 0.057, respectively).  The largest attractive BCS channel is shown 
in (D), but it has a very small coefficient (-0.020) and the rapid 
oscillations suggest that the calculation is not accurate 
at this point.  In the calculation, the Fermi surface was divided into $M = 24$ patches,
instead of just 16, to improve the accuracy in the higher wave channels.} 
\label{fig:tri_bcs}
\end{figure}

For the special case of the isotropic triangular lattice 
($t_1 = t_2 = 0.5$) at half-filling, the weak-coupling RG flows do not show any BCS 
instabilities.  The dominant BCS channels, 
$d_{x^2-y^2}$, $d_{xy}$ and $p$, are all repulsive.  These
channels are depicted in 
Fig. \ref{fig:tri_bcs}(A), \ref{fig:tri_bcs}(B) and \ref{fig:tri_bcs}(C). 
Fig. \ref{fig:tri_bcs}(D) shows the first attractive channel that develops, but 
the rapid oscillations in the effective potential, and its small size, indicate that the 
calculation is not reliable.  

\begin{figure}[!ht]
\includegraphics[width=3in,clip]{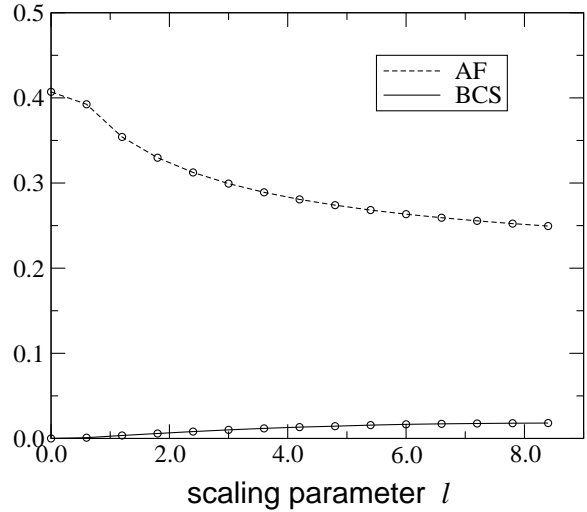}
\caption{Flow of dominant AF and BCS channels for the case of the 
isotropic triangular lattice.}
\label{fig:afxsc_tri}
\end{figure}

\begin{figure}[!ht]
\includegraphics[width=6in]{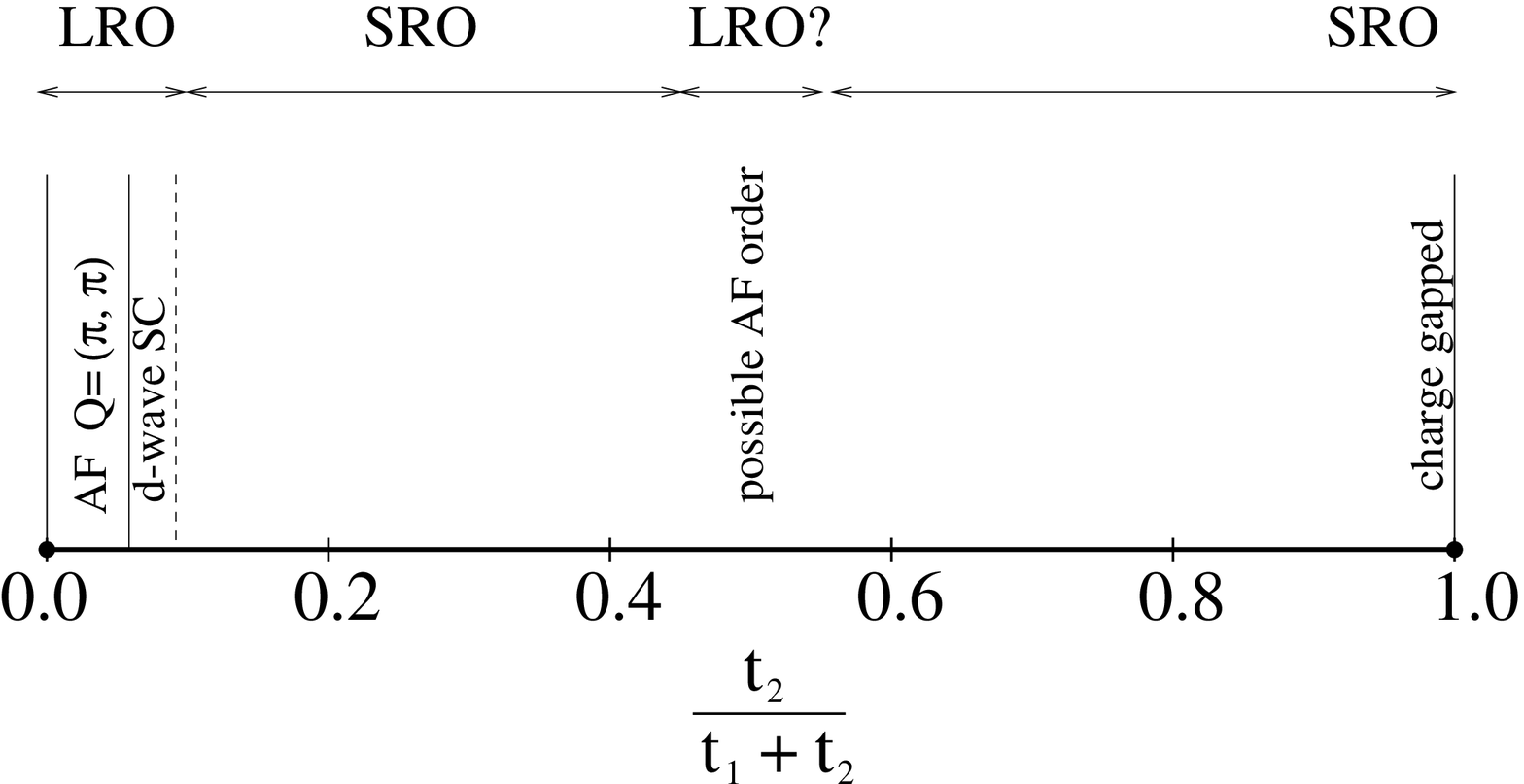}
\caption{Four regions of the $U > 0$ phase diagram, LRO / SRO / LRO / SRO, 
identified as $t_2/(t_1+t_2)$ varies 
from $0$ (square lattice) to $1$ (decoupled chains).}
\label{fig:phasediag}
\end{figure}

In Fig. \ref{fig:afxsc_tri} the dominant AF and BCS channels are compared.  Neither 
channel shows strong divergences, but the AF channel is significantly larger 
than the BCS channel.  Thus there are signs of re-entrant antiferromagnetic 
long-range order.  We speculate that there exist four different regions as
the isotropy parameter $t_2/(t_1+t_2)$ changes 
form $0$ (square lattice) to $1$ (decoupled chains).  These phases are shown in 
Fig. \ref{fig:phasediag}.  At $t_2 = 0$ the system exhibits long-range 
antiferromagnetic N\'eel order (LRO) with ordering vector $\vec{Q} = (\pi, \pi)$.  
But this long-range order is suppressed by turning on $t_2$.  Instead  
$d_{x^2-y^2}$ BCS instabilities dominate and only short-range order (SRO) occurs.  
Both AF and BCS instabilities are 
suppressed as $t_2$ is further increased and the nesting is completely 
eliminated.  Nevertheless, the AF coupling remains significantly larger 
than the BCS coupling.  
In the strong-coupling limit $U \rightarrow \infty$ the model can be mapped onto 
a nearest-neighbor spin-1/2 Heisenberg antiferromagnet which of course is insulating.  
On the isotropic triangular lattice this antiferromagnet
exhibits long-range AF order\cite{Singh,Bernu} with ordering wave vector 
$\vec{Q} = (4\pi/3, 0)$.  The question of whether or not our weak-coupling analysis
can describe this strong-coupling limit is tantamount to asking whether or not one
or more intermediate-coupling fixed points intervene between the repulsive weak-coupling
fixed point that is accessible in our RG analysis, and the attractive strong-coupling fixed point
which describes the antiferromagnetic insulator.   
Finally as $t_2$ becomes larger than $t_1$, the chains begin to decouple.  In the 
extreme limit of independent chains there can be no long-range spin order, as the 
Mermin-Wagner theorem tells us that continuous symmetries cannot break in 1+1 dimensions.  

Curiously, four such regions, LRO / SRO / LRO / SRO, have also 
been identified in the corresponding strong-coupling Heisenberg antiferromagnet 
with two exchange couplings $J_1$ and $J_2$.  The phase diagram of this model 
has been studied via a straightforward $1/S$ expansion\cite{Merino,Trumper}, a series 
expansion method\cite{Weihong}, and a large-$N$ Sp(N) solution\cite{Chung}.  All 
three methods find two regions of long-range order:  near the limit of a 
square lattice ($J_2 = 0$) and near the isotropic point ($J_2 = J_1$). It is 
remarkable that our weak-coupling RG analysis shows similar behavior.

The Hubbard model on the triangular lattice has been studied 
at intermediate values of the interaction strength 
within the Hartree-Fock approximation\cite{Krishnamurthy,Jayaprakash} 
and within the slave-boson 
method\cite{Gazza,Feiguin,Capone}.  A Mott-Hubbard metal-insulator transition 
is found to occur at a relatively large value of $U_c$ ($U_c = 5.27 t$ for 
the Hartree-Fock calculation\cite{Krishnamurthy} and $U_c = 7.23 t$ or 
$7.68 t$ from the slave-boson calculations\cite{Gazza,Capone}).  At a smaller value 
of $U = U_{c1}$ there is also a continuous transition from a paramagnetic 
metallic phase to a metallic phase with incommensurate spiral order.
The Hartree-Fock calculation yields $U_{c1} = 3.97 t$ and 
the slave-boson calculation gives $U_{c1} = 6.68 t$ for this transition.  
Signs of re-entrant AF order in our weak-coupling RG calculation, 
namely the relatively large size of the AF channels in comparison with the BCS channels, 
are broadly consistent with this picture, as AF tendencies can be a precursor
to a transition to the insulating state.

\section{Conclusion}
\label{sec:Conclusion}

Hubbard models have received extensive study in the 
context of high-$T_c$ superconductivity.  We have reproduced the well-known 
result that there is an AF instability at half-filling on the square lattice with
repulsive on-site Coulomb interaction and nearest-neighbor hopping.  Furthermore, upon
doping the system away from half filling, a crossover to a BCS regime with 
$d_{x^2-y^2}$ pairing symmetry occurs as expected.  We have shown that it is important
to retain subleading, formally irrelevant, corrections to the RG flows when the bare interaction
is repulsive and the Fermi surface is nearly nested.  We also studied 
another way of triggering a BCS instability.  Keeping the system 
at half filling, but introducing the diagonal hopping $t_2$ as shown in Fig. 
\ref{fig:tri} along one of the two diagonals, breaks up perfect nesting.  
Corresponding magnetic frustration kills the spin density wave, 
and Cooper pairing dominates, at least if $t_2$ is not too large.  
This result suggests that superconductivity can occur in a model of 
strongly correlated electrons, even at half-filling.
We emphasize that stripes are not expected to play a role here; 
even moderate on-site Coulomb repulsion should inhibit charge
segregation at half-filling.  

The half-filled Hubbard model on the anisotropic 
triangular lattice has been proposed as a minimal 
description of the conducting layers of the $\kappa$-(BEDT-TTF)$_2$X 
organic superconductors.  It is important to establish the pairing symmetry of
the superconducting state of these materials.  Our theoretical results predict
pairing of the $d_{x^2-y^2}$ type and in fact signs of such order have been seen
experimentally. 
We did not find any evidence of spontaneous time-reversal symmetry breaking. 
Pairing symmetry of the type $d_{x^2-y^2} \pm i d_{xy}$ would occur 
if an attractive $d_{xy}$ channel arose in addition to the $d_{x^2-y^2}$ channel.  
We find that attractive $d_{xy}$ channels neither occur when next-nearest-neighbor 
hopping $t^\prime$ is included on the square lattice, nor when non-zero hopping $t_2$ 
is turned on along one of the two diagonal directions.  

Finally, we made contact with 
previous work on Heisenberg antiferromagnets on the anisotropic triangular lattice. 
Our weak-coupling RG calculation shows AF tendencies in two separate regimes -- tendencies  
which seem to correspond with the two AF ordered phases found previously at large-$U$.  In particular
the portion of the phase diagram between the isotropic point ($t_1 = t_2$) and 
decoupled chains ($t_1 = 0$) is the relevant region for the layered Cs$_2$CuCl$_4$ antiferromagnet 
insulator material.  The competition we found between antiferromagnetic order and spin-liquid
behavior in our RG calculation may be consistent with the observed ease by which spin
order is destroyed in the compound\cite{Radu}.

\vskip 0.2in
{\bf Acknowledgments}
We thank Chung-Hou Chung, Tony Houghton, Ross McKenzie, and Matthias Vojta 
for useful discussions.  This work was supported in part by the NSF under 
Grant Nos. DMR-9712391 and DMR-0213818.

\end{document}